\documentclass[pre,twocolumn,superscriptaddress,showpacs]{revtex4-1}
\usepackage{graphicx}
\usepackage{amssymb}
\usepackage{amsmath}
\usepackage{bm}
\usepackage{color}
\newcommand{\blue}[1]{\textcolor{black}{#1}}

\newcommand{\figref}[1]{Fig.~\ref{#1}}

\newcommand{\bitem}{\begin{itemize}}
\newcommand{\eitem}{\end{itemize}}
\newcommand{\benum}{\begin{enumerate}}
\newcommand{\eenum}{\end{enumerate}}
\newcommand{\btab}[1]{\begin{tabular}{#1}}
\newcommand{\etab}{\end{tabular}}
\newcommand{\btabn}[1]{\begin{tabular}{#1}}
\newcommand{\etabn}{\end{tabular}}
\newcommand{\beq}{\begin{equation}}
\newcommand{\eeq}{\end{equation}}
\newcommand{\beqn}{\begin{equation*}}
\newcommand{\eeqn}{\end{equation*}}
\newcommand{\bsplit}{\begin{split}}
\newcommand{\esplit}{\end{split}}

\newcommand{\kB}{\ensuremath{k_{\mathrm{B}}}}
\newcommand{\phig}{\ensuremath{\phi_{\mathrm{g}}}}

\newcommand{\hide}[1]{}
\newcommand{\gdot}{\dot\gamma}

\newcommand{\rr}{\mathbf{r}}
\newcommand{\xx}{\mathbf{x}}

 \definecolor{green4}{rgb}{0,0.6,0}
 \definecolor{blue4}{rgb}{0,0,0.6}
\definecolor{gray4}{rgb}{0.6,0.6,0.6}

\newcommand{\delete}[1]{}

\begin{document}

\title{Correlations of plasticity in sheared glasses}

\author{Fathollah Varnik}
\email{fathollah.varnik@rub.de}
\affiliation{Max-Planck Institut f\"ur Eisenforschung, Max-Planck Str.~1, 40237 D\"usseldorf, Germany}
\affiliation{Interdisciplinary Centre for Advanced Materials Simulation (ICAMS), Ruhr-Universit\"at Bochum, Stiepeler Strasse 129, 44801 Bochum, Germany}
\author{Suvendu Mandal}
\affiliation{Max-Planck Institut f\"ur Eisenforschung, Max-Planck Str.~1, 40237 D\"usseldorf, Germany}
\author{Vijaykumar Chikkadi}
\affiliation{Institute of Physics, University of Amsterdam, Science Park 904, 1098 XH Amsterdam, The Netherlands}
\author{Dmitry Denisov}
\affiliation{Institute of Physics, University of Amsterdam, Science Park 904, 1098 XH Amsterdam, The Netherlands}
\author{Peter Olsson}
\affiliation{Department of Theoretical Physics, Ume\r{a} University, 901~87 Ume\aa, Sweden}
\author{Daniel V\r{a}gberg}
\affiliation{Department of Theoretical Physics, Ume\r{a} University, 901~87 Ume\aa, Sweden}
\author{Dierk Raabe}
\affiliation{Max-Planck Institut f\"ur Eisenforschung, Max-Planck Str.~1, 40237 D\"usseldorf, Germany}
\author{Peter Schall}
\affiliation{Institute of Physics, University of Amsterdam, Science Park 904, 1098 XH Amsterdam, The Netherlands}

\begin{abstract}
In a recent paper [S. Mandal \emph{et al.}, Phys.\ Rev.\ E {\bf 88}, 022129 (2013)] the nature of spatial correlations of plasticity in hard sphere glasses was addressed both via computer simulations and in experiments. It was found that the experimentally obtained correlations obey a power law whereas the correlations from simulations are better fitted by an exponential decay. We here provide direct evidence--- via simulations of a hard sphere glass in 2D---that this discrepancy is a consequence of the finite system size in the 3D simulations. By extending the study to a 2D soft disk model at zero temperature [D.\ J.\ Durian, Phys.\ Rev.\ Lett.\ {\bf 75}, 4780 (1995)], the robustness of the power-law decay in sheared amorphous solids is underlined. Deviations from a power law occur when either reducing the packing fraction towards the supercooled regime in the case of hard spheres or changing the dissipation mechanism from contact dissipation to a mean-field type drag for the case of soft disks.
\end{abstract}

\date{\today}
\maketitle
\section{Introduction}
It is well established that dynamic correlations grow in glass forming liquids as the glass transition is approached \cite{Bennemann1999a,Scheidler2002,Baschnagel2005,Ballesta2008,Varnik2009,Berthier2011a,Kob2012}. This growth is, however, generally found to be limited to a few particle diameters~\cite{Scheidler2002,Ballesta2008,Varnik2009,Kob2012}. A dramatic change may occur if the glass is driven by an applied shear that forces structural rearrangements~\cite{Chikkadi2011}; such external driving can lead to avalanche-like plastic response, mediated by a long range elastic field~\cite{Lemaitre2009}.

Some of us recently addressed this issue for a hard sphere glass both via computer simulations and experiments with a focus on direction-dependence of correlations and the crossover from the thermal regime of supercooled liquids to the athermal limit of strongly driven glasses~\cite{Chikkadi2012b,Mandal2013}. Qualitative agreement was found between simulations and experiments regarding both the behavior of single particle fluctuations (found to be isotropic) and the anisotropy of their spatial correlations. The specific functional form of these correlations was, however, found to be different. While experimental data were best described by a power law decay---recalling a self-similar behavior---simulations suggested an exponential decay with a characteristic length of the order of a few particle diameters.

Here we address this issue and provide strong evidence that a reason for this discrepancy is the finite system size of simulations. When performing event-driven finite-temperature simulations in 2D (which allows for much larger linear sizes, $L$), we find that the exponential decay found for smaller sizes changes to an algebraic decay at larger $L$.

To plunge more deeply into the question of exponentially and algebraically decaying correlations, we also perform simulations of a Durian-type 2D soft disk model at zero temperature~\cite{Durian1995} with two different types of hydrodynamic drag: (i) contact dissipation (CD), where the dissipative force is proportional to the relative velocity of interacting particle pairs and (ii) mean-field or reservoir dissipation (RD), where a drag force relative to an externally imposed background linear velocity field is used (see Sec.~\ref{sec:models} for details). These simulation studies are accompanied by experiments on granular particles. Results obtained for the CD-case are in line with experiments, thus showing that power-law correlations are the generic response of driven amorphous solids.

In the case of reservoir dissipation, on the other hand, there are strong deviations from a power-law decay even for the largest $L$ simulated. We attribute this behavior to the simplified dissipation mechanism which couples the particle dynamics to an externally imposed flow, without any influence of the particle motion on the flow velocity \cite{Tighe:2010,Woldhuis2013,Vagberg-cdrd}.

\section{Simulation model and experimental system} \label{sec:simulation-setup}
\label{sec:models}

We perform two kinds of simulations: Our first simulation model is a polydisperse hard sphere system of mass $m=1$ and average diameter $\sigma=0.8$. Lengths are measured in units of $\sigma$ and time in units of $\sigma \sqrt{m/\kB T}$, where $T$ is temperature and $\kB$, the Boltzmann constant, both set to unity for convenience. Event-driven Molecular Dynamics simulations are performed using the Dynamo code \cite{DynamO}. Periodic boundary conditions are used along all directions. When combined with the Lees-Edwards boundary condition \cite{Evans1986}, this leads to a shear deformation of $\gamma=t\dot \gamma$. The shear rate $\dot \gamma$ varies around $10^{-4}$. The packing fractions studied are around the glass transition point, which, for the present polydisperse system, is located at a packing fraction of $\phig \approx 0.58$ (3D) \cite{Pussey2009} and $\phig \approx 0.80$ (2D) \cite{Santen2000,Bayer2007}. The quiescent properties of the 3D system have been studied extensively in Ref.~\onlinecite{Williams2001}. The temperature is fixed at $T=1$ via velocity rescaling. We present all the measurements after $100\%$ shearing to ensure that the system has reached steady state.

We also perform simulations with soft bidisperse particles with size ratio 1.4 in two
dimensions and at zero temperature. The diameter of the smaller particles is
$\sigma=1$. With $\sigma_i$ for the diameter of particle $i$ the mass is
$m_i=\pi\sigma_i^2/2$. We make use of Lees-Edwards boundary conditions with shear rates
$\gdot=10^{-6}$ through $10^{-4}$. At the densities considered here, there is no strong
dependence on shear rate, and the data in this paper is only for a single shear rate,
$\gdot=10^{-5}$. The number of particles in the simulations is mostly $N=262144$, though
smaller systems are also simulated to examine finite size effects.
In the model there is both a conservative elastic force and a dissipative force. The elastic force depends on the position coordinates only. With $\rr_{ij} = \rr_i - \rr_j$ for the distance between particles $i$ and $j$, $\sigma_{ij} = (\sigma_i+\sigma_j)/2$, and the overlap $\delta_{ij} = (\sigma_{ij}-r_{ij})/\sigma_{ij}$, the elastic force is $f^\mathrm{el}=k_e \delta_{ij} \hat\rr_{ij}$ \blue{with $\hat\rr_{ij}=\rr_{ij}/|\rr_{ij}|$}. We note that at shear rate $\gdot=10^{-5}$ the particles deform only marginally, $\delta_{ij}\approx 1.1\times 10^{-4}$ at $\phi=0.82$. This means that we expect this soft disk model to behave essentially as a hard disk model; the softness of the particles should not play any significant role.  We report on results obtained both with contact dissipation (CD) for which the dissipative force is given by the velocity difference to all particles in contact, $f^\mathrm{diss}_\mathrm{CD}= -k_d\sum_j (\mathbf{v}_i - \mathbf{v}_j)$, as well as with reservoir dissipation (RD), $f^\mathrm{diss}_\mathrm{RD}= -k_d (\mathbf{v}_i - \mathbf{v}_{\text{R}}(\rr_i))$, where $\mathbf{v}_{\text{R}}(\rr_i) = \gdot y_i\hat{\xx}$ \blue{(with $\hat{\xx}$ the unit vector along the $x$-direction)}. The latter model has sometimes been called the mean-field model, as the particles dissipate against the average velocity $\gdot y_i\hat{x}$. In the simulations we take $k_d=1$ and $k_e=1$.

For the experimental measurements, we use suspensions of both Brownian and granular particles. The Brownian system consists of sterically stabilized polymethylmethacrylate (PMMA) particles with diameter of $\sigma = 1.3 \mu$m, suspended in a mixture of Cycloheptyl Bromide and Cis-Decalin. This solvent mixture matches both the density and refractive index of the particles. The particles have a polydispersity of $ 7\%$ to prevent crystallization. The particle volume fraction is fixed at $\phi \sim 0.6$, well inside the glassy state~\cite{vanMegen1998}. We apply shear at constant rate in the range of $1.5 \times 10^{-5}$ to $2.2 \times 10^{-4} s^{-1}$, corresponding to modified Peclet numbers $\dot{\gamma} \tau$ between 0.3 and 2.2, respectively~\cite{Chikkadi2011}. Here, the structural relaxation time is $\tau = 2 \times 10^4 $ s, as determined from the mean-square displacement of the particles. The granular system consists of PMMA particles with a diameter of 3.9 mm, and a polydispersity of $\sim 5 \%$ suspended in mixture of dimethyl-sulfoxide, water and salt (NaI), carefully tuned to match both the refractive index and the density of the particles~\cite{Lorincz2010}. Individual particles are imaged in three dimensions using a laser sheet, while the suspension is sheared at a rate of $\sim 5 \times 10^{-5}$s$^{-1}$ with a constant confining pressure of $\sim 7$ kPa ~\cite{Lorincz2010}.

\section{Spatial correlations of plasticity}

\blue{We define a quantitative measure of plastic activity as follows \cite{Falk1998}. For a reference particle (noted here with index 0), we follow the evolution of the distance vectors ${\bf d}_i={\bf r}_i-{\bf r}_0$ for a short time interval $\delta t$, where $i$ runs over the nearest neighbors of the reference particle. We then define a measure for plastic activity as $D^2 = (1/n) {\sum_{i=1}^{n}}({\bf d}_i(t + \delta t)-{\bf d}_i(t) - \bm{\epsilon}\cdot {\bf d}_i(t))^2$, where $\bm{\epsilon}$ is the linear (affine) transformation tensor which best describes the time evolution of ${\bf d}_i$. $D^2$ is the mean-square deviation from a local affine deformation, and is known as an excellent  measure of local plasticity \cite{Falk1998}.}

\blue{The above applies directly to both the experiments and the event-driven (finite temperature) simulations. The analyses in the zero-temperature simulations (the RD and CD models) work the same except that ${\bf d}_i(t + \delta t)-{\bf d}_i(t)$ is replaced by ${\bf v}_i(t)\delta t$, where ${\bf v}_i(t)$ is the instantaneous velocity of the particle $i$ relative to the reference particle. Note that, here, $\delta t$ affects the magnitude of the optimum affine transformation tensor (and thus $D^2$) by a constant factor only and has no effect on the spatial correlations of $D^2$ addressed in this paper.}

We then use the above introduced scalar quantity $D^2$ to define the correlation function~\cite{Chikkadi2011}
\begin{equation}
  C_{D^2}({\Delta \mathbf{r}}) = \frac{ \left< D^2({\bf r} + \Delta \mathbf{r}) D^2(\mathbf{r})
    \right> - \left< D^2(\mathbf{r}) \right> ^{2} } { \left< D^2(\mathbf{r})^{2}
    \right> - \left< D^2(\mathbf{r}) \right> ^{2} }.
  \label{c_r}
\end{equation}
The function $C_{D^2}(\Delta \mathbf{r})$ provides a measure of correlations between plastic activity at two points in space separated by a vector $\Delta {\mathbf r}$.

\section{Results on spatial correlations}

\subsection{Motivation---experiments and event-driven simulations in 3D}
\label{sec:background}

\begin{figure}
  \unitlength=1mm
  \begin{picture}(0,0)(0,3) \put(0,0.5){(a)} \put(50,0.5){(b)} \put(0,-35){(c)}
    \put(50,-35){(d)}
  \end{picture}
  \includegraphics*[width=0.24\textwidth]{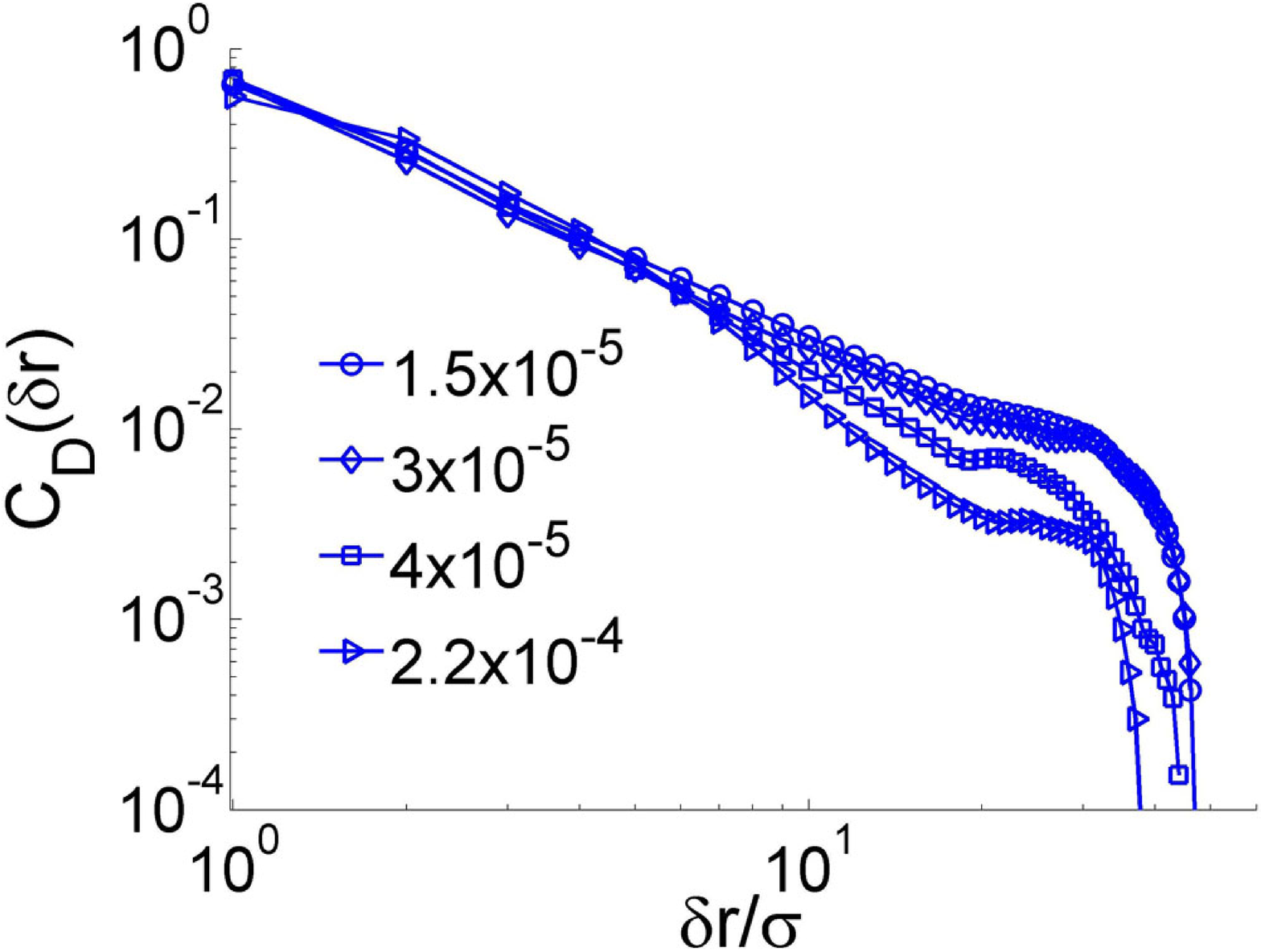}\hfill
  \includegraphics*[width=0.22\textwidth]{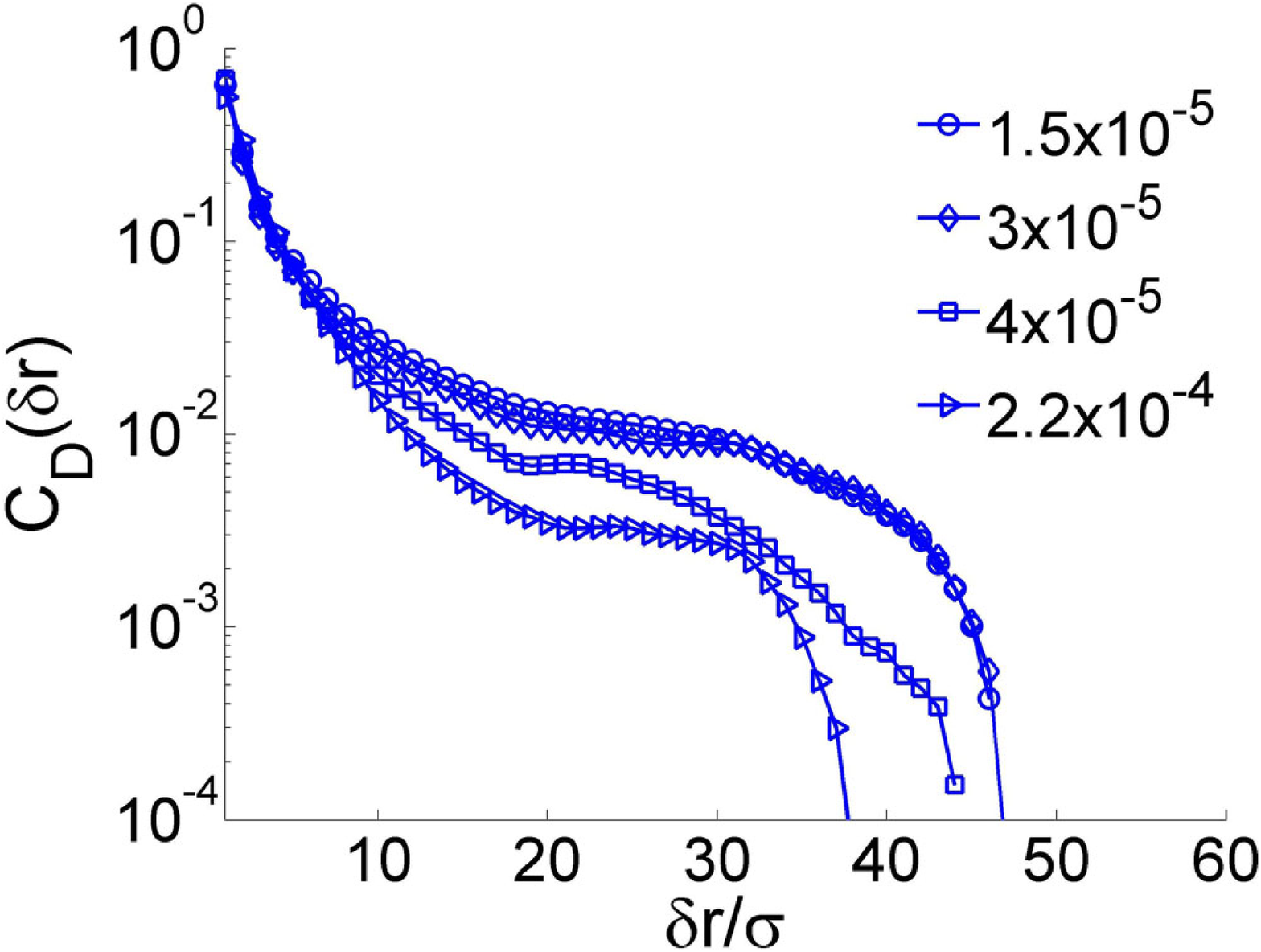}\vspace*{5mm}
  \includegraphics*[width=0.23\textwidth]{Simulations3D_loglog.eps}\hfill
  \includegraphics*[width=0.23\textwidth]{Simulations3D_loglin.eps}\vspace*{2mm}
  \caption[]{Correlation of plastic activity in experiments (a,b) and simulations
    (c,d) of hard-sphere glasses. The same data are shown both in double logarithmic (left
    panel) and semilogarithmic (right panel) scales. The experimental data show robust
    power-law decay while the simulations rather show exponential decay.}
\label{fig:CD2-expsim}
\end{figure}

We compare correlations of plastic activity in experiments and the 3D simulations in \figref{fig:CD2-expsim}. Panels (a) and (b), which are the experimental correlations, show robust power-law correlations with exponent $\alpha\sim1.3$ for a range of Peclet numbers. These power-law correlations extend out to the distance $r \sim 50 \sigma$ that equals the vertical system size.  (In the horizontal directions, the experimental system is macroscopically large.) In contrast, panels (c) and (d) of \figref{fig:CD2-expsim}, which are the correlations from the event-driven 3D simulations, give strong evidence of an exponential decay. This discrepancy was noted by some of us in an earlier publication~\cite{Mandal2013}.

\subsection{Event-driven simulations in 2D}
\label{sec:res:hard}

\begin{figure}
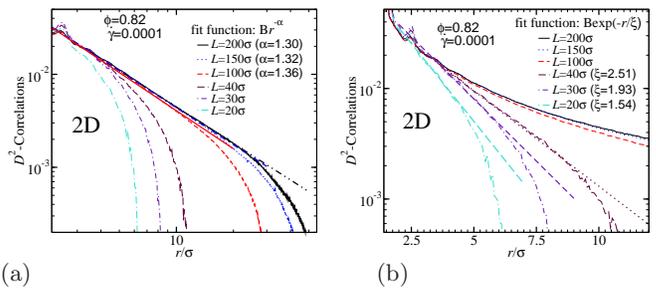

\unitlength=1mm
\begin{picture}(0,0) (0,3)
\put(0,0){(a)} \put(50,0){(b)}
 \end{picture}
  \includegraphics*[width=0.23\textwidth]{Simulations2D_phi82_loglog.eps}\hfill
  \includegraphics*[width=0.23\textwidth]{Simulations2D_phi82_loglin.eps}\vspace*{2mm}
  \caption[]{Angle averaged correlation of plastic activity in 2D simulations of hard disks for
    various system sizes. The packing fraction is $\phi=0.82>\phig\approx 0.80$  (glassy phase).
    The same data are shown both in double logarithmic (a) and
    semilogarithmic scale (b).
}
  \label{fig:CD2-FV}
\end{figure}

One possible reason for the discrepancy noted above could be the limited system size, $L/\sigma=25$, in the event-driven 3D simulations which is due to limitations in computational resources. To achieve larger linear system sizes with similar computational effort we turn to two dimensions. In this way, we are able to reach sizes of up to $L/\sigma=200$. We thus performed a systematic study of the finite size effects of $C_{D^2}$ at a density of $\phi=0.82$ \blue{above $\phig \approx 0.80$} (\figref{fig:CD2-FV}). For $L/\sigma<50$, we find that correlations decay exponentially, in perfect agreement with the 3D simulations in \figref{fig:CD2-expsim}(c) and (d). For larger sizes, the figure shows clear evidence of a power-law decay. This observation suggests that it is the rather limited system size in the 3D simulations that is the main cause for the observed exponential decay, and that a power-law decay is the true behavior for sufficiently large system sizes. These results therefore point to a good agreement between experiments and simulations.

\begin{figure}
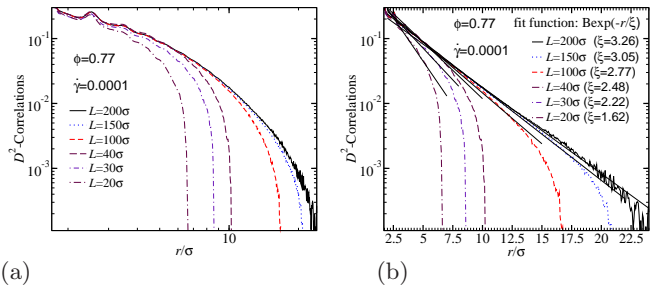

  \unitlength=1mm
  \begin{picture}(0,0) (0,3)
\put(0,0){(a)} \put(50,0){(b)}
  \end{picture}
  \includegraphics*[width=0.23\textwidth]{Simulations2D_phi77_loglog.eps}\hfill
  \includegraphics*[width=0.23\textwidth]{Simulations2D_phi77_loglin.eps}\vspace*{2mm}
  \caption[]{Correlation of plastic activity in 2D simulations of hard disks for
    various system sizes as indicated. The packing fraction is $\phi=0.77<\phig\approx 0.80$
    (supercooled state). The same data are shown both in double logarithmic (a) and
   semilogarithmic scale (b). 
}
\label{fig:CD2-FV2}
\end{figure}

We have also performed simulations at a lower packing fraction of $\phi=0.77$ (supercooled state). In contrast to the algebraic decay in the glassy phase, $C_{D^2}$ now decays exponentially at the largest system size investigated, as clearly shown in Fig.~\ref{fig:CD2-FV2}. This observation is interesting and suggests that the glassy state is clearly distinct from the supercooled state: spatial correlations of plastic activity are mediated by the elastic field \cite{Picard2005,Schall2007,Lemaitre2009}. In the supercooled state, the elasticity is not well established (though observable for sufficiently fast processes~\cite{Chattoraj2013}), and correlations are short ranged; in the glassy phase, on the other hand, the glass has developed elasticity resulting in power-law correlations that are related to a system-spanning elastic field.

\subsection{2D simulations of soft particles at zero temperature}
\label{sec:res:soft}

\begin{figure}
\includegraphics*[width=0.238\textwidth, bb=41 328 390 580]{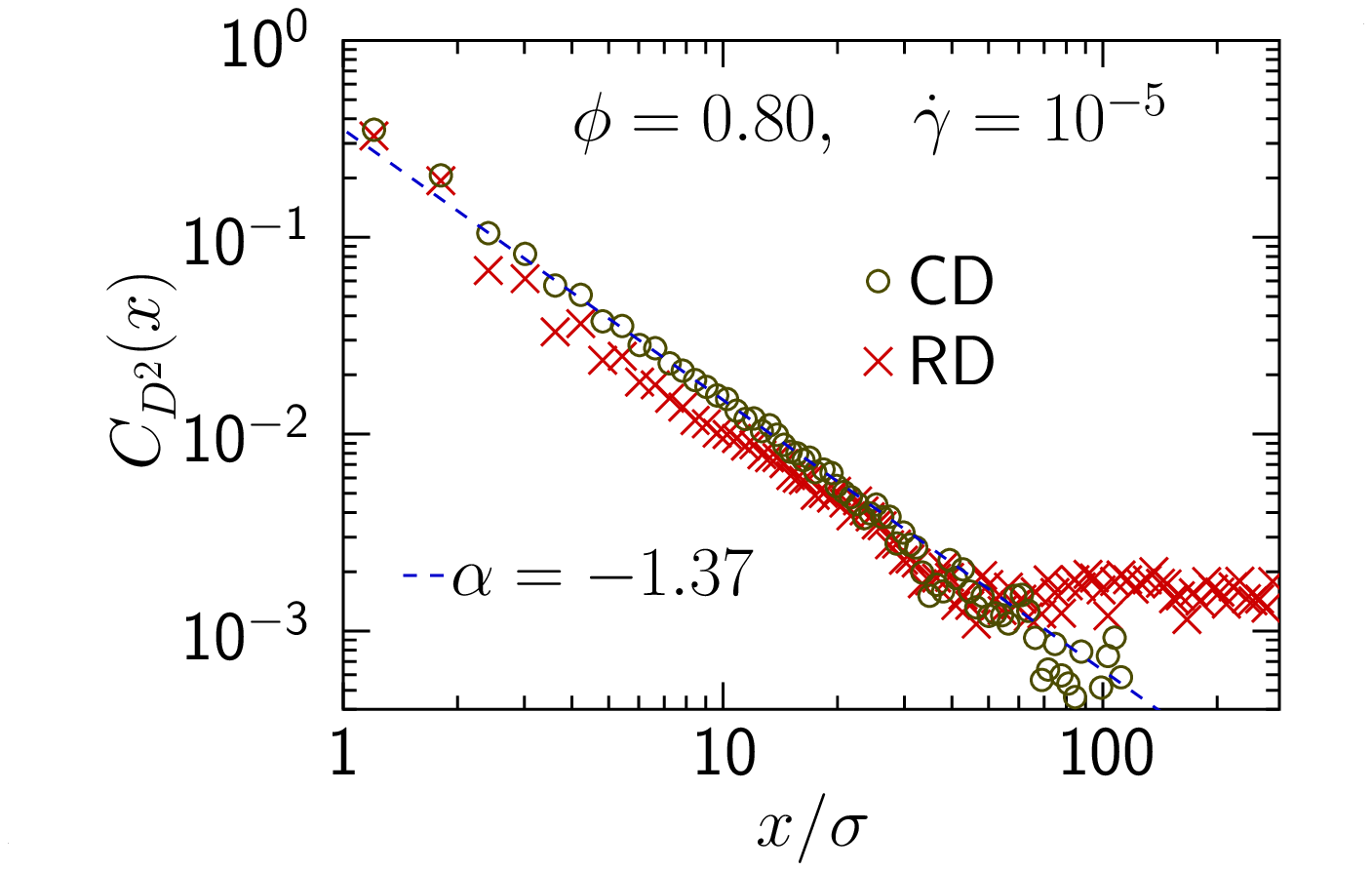}
\includegraphics*[width=0.238\textwidth, bb=41 328 390 580]{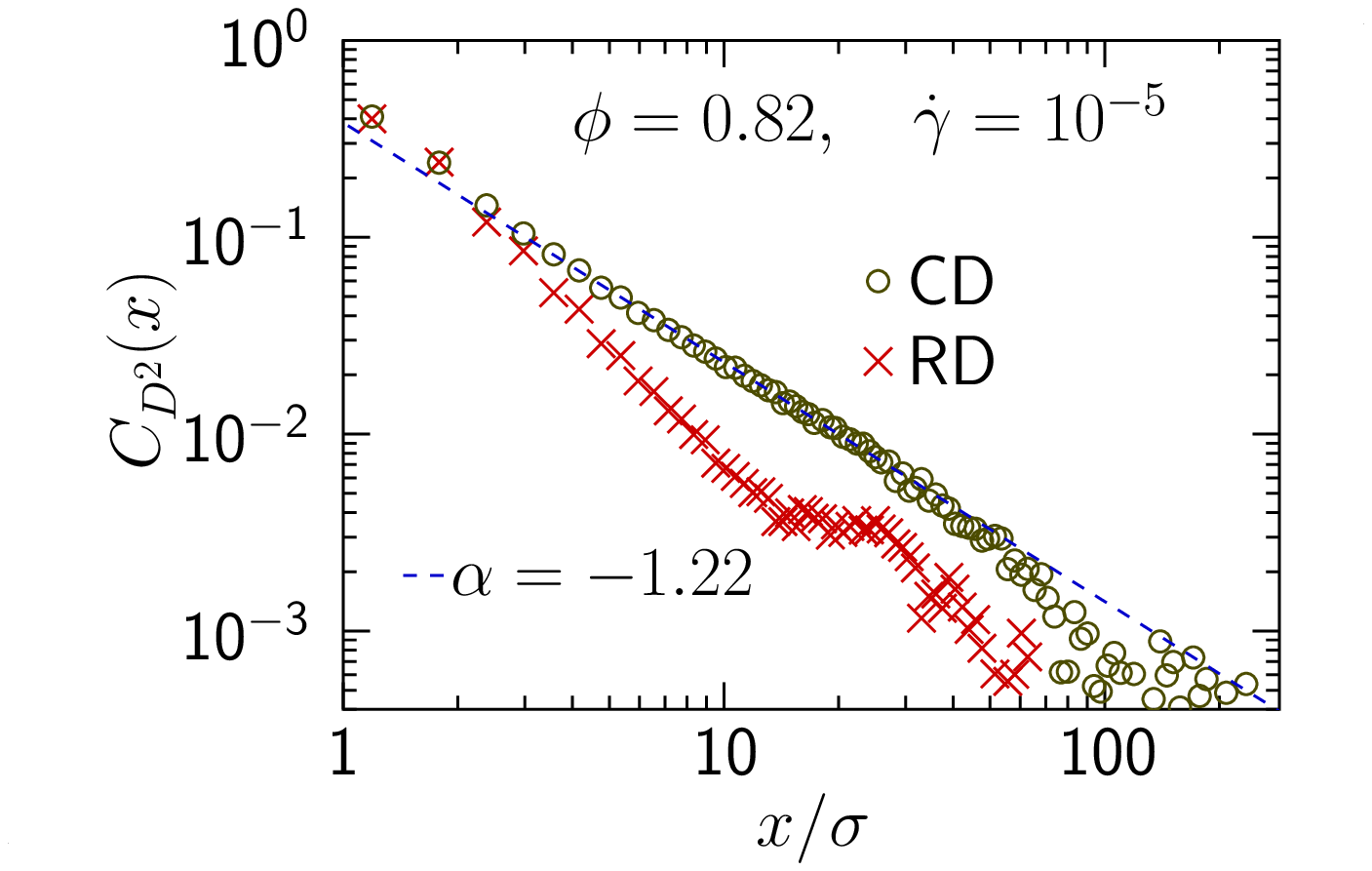}\\\vspace*{2mm}
\includegraphics*[width=0.238\textwidth, bb=41 328 390 580]{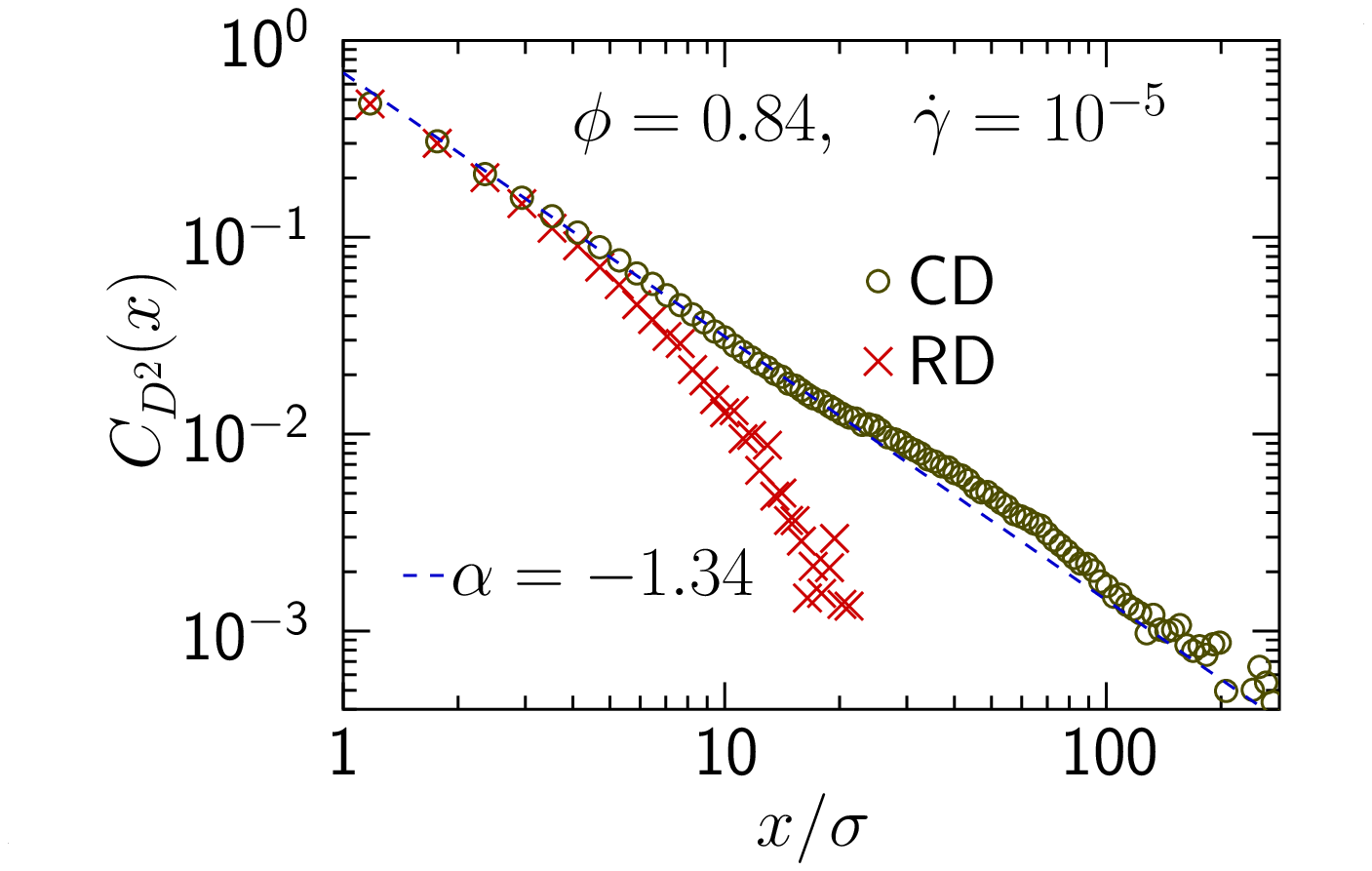}
\includegraphics*[width=0.238\textwidth, bb=41 328 390 580]{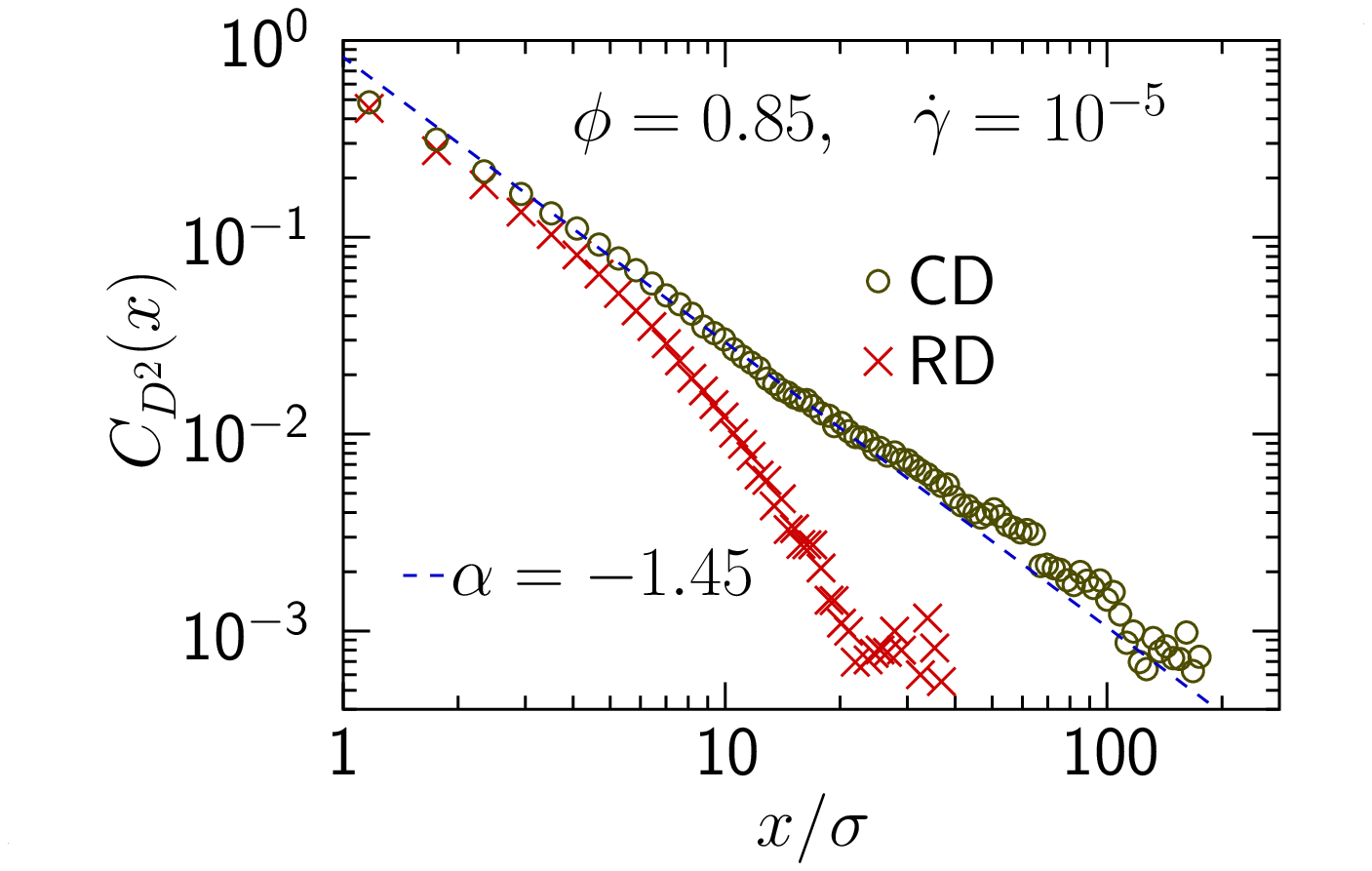}\\
\caption[]{Correlation of plastic activity along the flow direction from simulations of a
  Durian-type model with contact dissipation (CD) and reservoir dissipation (RD) for
  different densities across the jamming transition ($\phi_J\approx 0.843$).  The dashed
  lines and quoted exponents are obtained by fitting $C_{D^2}$ for the CD model to an
  algebraic decay for $2\leq x/\sigma\leq20$.  }
\label{fig:CD2-PO}
\unitlength=1mm
\begin{picture}(0,0)(50,-64)
\put(8,0){(a)} \put(52,0){(b)}
\put(8,-33.5){(c)} \put(52,-33.5){(d)}
\end{picture}
\end{figure}

\begin{figure}
\includegraphics*[width=0.238\textwidth, bb=41 328 390 580]{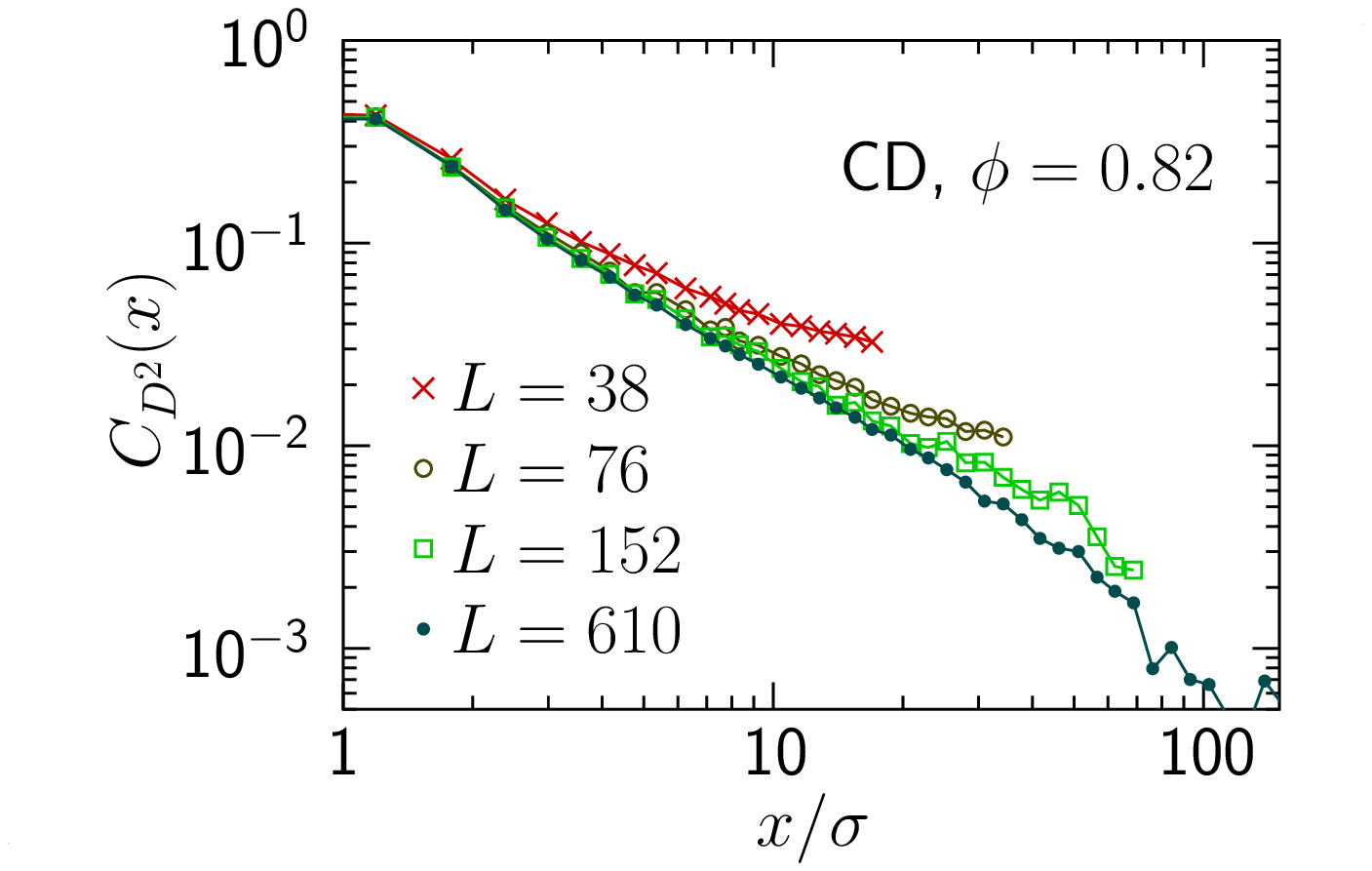}
\includegraphics*[width=0.23\textwidth]{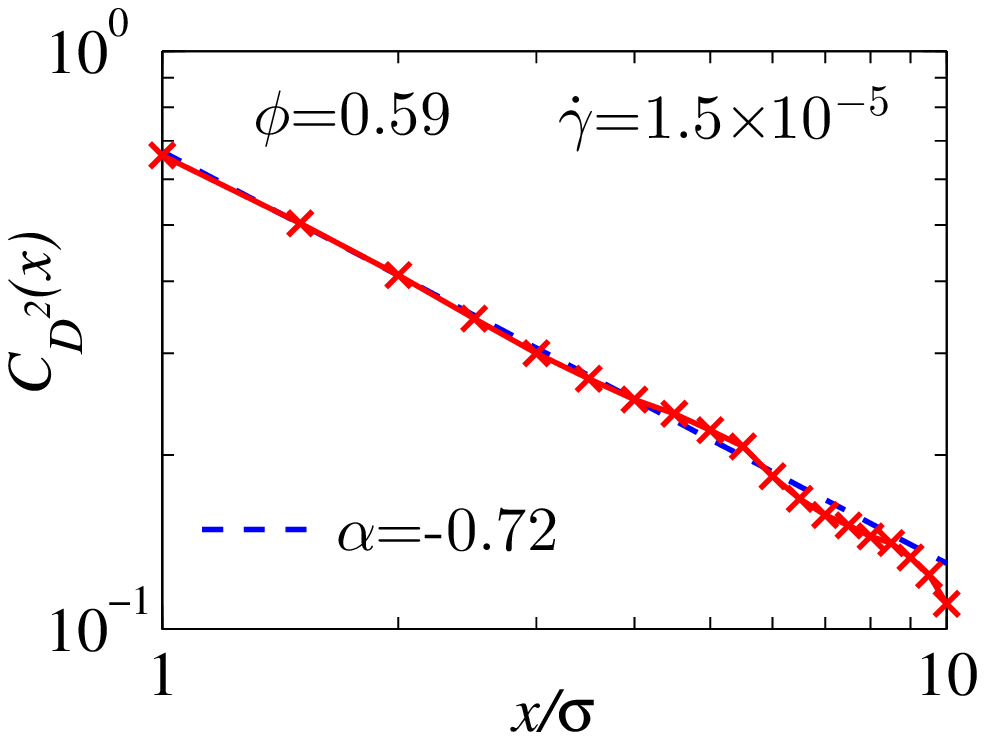}
\begin{picture}(0,0) (236,5)
\put(0,0){(a)}
\put(120,0){(b)}
\end{picture}\\
\caption{(a) Correlation of plastic activity along the flow direction from simulations of
  a Durian-type model with contact dissipation for different system sizes. (b)
  Experimental result for a granular suspension.
}
\label{fig:CD2-PO+PS}
\end{figure}

Since the particles in the experimental system experience drag forces when they pass by one another, but the simulated particles do not, it is interesting to try and examine the role of drag forces for the occurrence of algebraically decaying correlations. To this end, we have performed 2D simulations of a Durian-type model of \blue{athermal} soft disks for both CD and RD as described in Sec.~\ref{sec:models}. The two models are compared in Fig.~\ref{fig:CD2-PO}, where $C_{D^2}$ along the flow direction is shown at different densities across the jamming transition. The algebraic behavior for the CD model is very robust and changes only weakly with density. The exponent $\alpha$ varies slightly with $\phi$ and remains close to $-1.3$. This value coincides with both the one obtained in the 2D hard-sphere simulations and the experiments. Interestingly, the RD model exhibits a different behavior: while the decay is essentially algebraic at the lowest density, $\phi=0.80$, it is rather exponential at higher densities, $\phi\gtrsim 0.84$. These results indicate that the dissipation mechanism plays a significant role in the correlations, as also indicated by recent works \cite{Tighe:2010,Woldhuis2013,Vagberg-cdrd}. The results thus indicate certain pitfalls in simulation models that one has to be aware of: as the RD model is a simplified model to describe dissipation, we conclude that the CD model better describes the real experimental situation, and that power-law correlations are the generic response of athermal driven suspensions. This is indeed confirmed in experiments by direct imaging of correlations in a sheared granular suspension: As shown in Fig.~\ref{fig:CD2-PO+PS}, power-law correlations are observed in the shear direction, in agreement with the CD model, provided the system size is sufficiently large. In the experiment, correlations can only be imaged for distances $r \lesssim 10d$ due to the small size of the experimental system that can be imaged in 3D. Nevertheless, \blue{despite} the limited system size, the power law becomes apparent, and the agreement between the experiments and the CD model becomes clear.\\

\section{Conclusion}
\label{sec:conclusion}

In this work, we address the nature of spatial correlations of plasticity in sheared amorphous solids in experiments and simulations using different simulation models. This study is motivated by a recent publication by a part of the present authors \cite{Mandal2013} reporting a power law decay in experiments on a colloidal hard sphere glass but an exponential decay in event-driven molecular dynamics simulations of a model hard sphere glass (\figref{fig:CD2-expsim}). By going to 2D that allows for simulations at much larger linear system sizes, we find strong evidence that the exponential behavior reported in Ref.~\onlinecite{Mandal2013} is a finite size effect (\figref{fig:CD2-FV}). This conclusion is underlined by an algebraic decay of correlations in Durian-type soft disks~\cite{Durian1995} at zero temperature (\figref{fig:CD2-PO+PS}). Furthermore, by comparing two simulation models, reservoir dissipation and contact dissipation, we find that hydrodynamic drag can affect the correlations: mean-field type drag forces that couple the dynamics of particles to an externally imposed flow field without any feed-back mechanism may strongly bias the nature of correlations (\figref{fig:CD2-PO}). Thus, while this work demonstrates the robust and genuine algebraic nature of correlations in flowing amorphous solids, it also \blue{sheds some light onto} the origin of possible deviations from power-law decays.

\section{acknowledgments}
S.M. is financially supported by the Max-Planck Society. P.S. acknowledges support by a VIDI fellowship from the Netherlands Organization for Scientific Research (NWO). ICAMS acknowledges funding from its industrial sponsors, the state of North-Rhine Westphalia and the European Commission in the framework of the European Regional Development Fund (ERDF).  P.O. and D.V. acknowledge support by the Swedish Research Council, Grant No.\ 2010-3725 and computational resources provided by the Swedish National Infrastructure for Computing (SNIC) at PDC and HPC2N.

\bibliographystyle{prsty}
\bibliography{literature-suvendu-mandal-v4}

\end{document}